\documentclass{PoS}

\title{Longitudinal and transverse spin transfer to $\Lambda$ and $\bar\Lambda$ hyperons in p+p collisions at STAR}

\ShortTitle{Spin transfer to $\Lambda$ and $\bar\Lambda$ hyperons at STAR}

\author{\speaker{Qing-hua Xu} , for the STAR Collaboration\\ 
       Key Laboratory of Particle Physics and Particle Irradiation (MoE), \\
       Institute of Frontier and Interdisciplinary Science,\\
       Shandong University, Qingdao, Shandong 266237, China\\
        E-mail: \email{xuqh@sdu.edu.cn}}

\abstract{
The longitudinal or transverse spin transfer to Lambda and anti-Lambda hyperons in polarized proton-proton collisions is expected to be sensitive to the helicity or transversity distributions of strange and anti-strange quarks of the proton, and to the corresponding polarized fragmentation function. We report the first measurement of the transverse spin transfer to $\Lambda$ and $\bar \Lambda$ along the polarization direction of the fragmenting quark, $D_{TT}$, in transversely polarized proton-proton collisions at 200 GeV with the STAR experiment at RHIC. The data correspond to an integrated luminosity of 18 pb$^{-1}$, and cover a kinematic range of |$\eta$|< 1.2 and transverse momentum $p_T$ up to 8 GeV/c. We also report an improved measurement of the longitudinal spin transfer $D_{LL}$ to $\Lambda$ and $\bar \Lambda$ with $p_T$ up to 6 GeV/c, using data with about twelve times larger figure-of-merit than the previously published STAR results. The prospects of hyperon polarization measurements in the forward pseudo-rapidity region (2.5<$\eta$<4) in p+p collision in the year of 2021 and beyond will also be discussed, which is based on the STAR forward detector upgrade plan including a forward tracking system and a forward calorimeter system.
}

\FullConference{23rd International Spin Physics Symposium - SPIN2018 -\\
		10-14 September, 2018\\
		Ferrara, Italy}

\begin{document}

\section{Introduction}

The polarizations of $\Lambda$ and $\bar\Lambda$ hyperons have been studied extensively in various aspects of spin effects in high energy reactions due to the self spin-analyzing parity violating decay.
In particular, $\Lambda$ polarization transferred from polarized lepton or hadron beams (usually referred to as ``spin transfer") in different reactions provides a natural connection to polarized fragmentation functions and the polarized parton densities of the nucleon. 
A number of measurements have been made in polarized lepton-nucleon deep inelastic scattering (DIS), and in polarized hadron-hadron collisions in the STAR experiment at RHIC.
As the $s(\bar{s})$ quark plays a dominant role in $\Lambda\,(\bar{\Lambda})$ hyperon's spin content, the measurements of spin transfer to $\Lambda\,(\bar{\Lambda})$ hyperon provide a natural connection to the polarized parton distribution of (anti-)strange quarks in nucleon, which are not well constrained yet in experiment.  
In particular, the transverse spin transfer to $\Lambda$ ($\bar\Lambda$) in hadron-hadron collisions provides a natural connection to transversity of (anti-)strange quarks through polarized fragmentation functions\,\cite{deFlorian:1998am,Xu:2004es,Xu:2005ru}. 
Similarily, the longitudinal spin transfer to $\Lambda$ ($\bar\Lambda$) hyperons in hadron-hadron collisions provides sensitivity to the helicity distribution of (anti-)strange quarks~\cite{deFlorian:1998ba,Boros:2000ya,Ma:2001na,Xu:2002hz,Xu:2005ru,Chen:2007tm}.

In this contribution, we report the first measurement of transverse spin transfer to $\Lambda$ and $\bar\Lambda$ hyperons in transversely polarized proton-proton collisions at $\sqrt s=200$\,GeV with the STAR experiment at RHIC~\cite{Adam:2018dtt}. 
Then we report an improved STAR measurement of the longitudinal spin transfer $D_{LL}$ to $\Lambda$ and ${\bar \Lambda}$ hyperons in longitudinally polarized proton--proton collisions at $\sqrt{s}$ = 200\,GeV\,\cite{Adam:2018kzl}.

\section{First measurement of transverse spin transfer to $\Lambda$ and $\bar\Lambda$ hyperons at RHIC}

The transverse spin transfer, $D_\mathrm{TT}$, to the $\Lambda$ in proton-proton collisions is defined as:
\begin{equation}
  \centering
  D_\mathrm{TT}
  \equiv
  \frac{d\sigma^{{(p^{\uparrow}p \rightarrow \Lambda^{\uparrow}X)}}-d\sigma^{{(p^{\uparrow}p \rightarrow \Lambda^{\downarrow}X)}}} {d\sigma^{{(p^{\uparrow}p \rightarrow \Lambda^{\uparrow}X)}}+d\sigma^{{(p^{\uparrow}p \rightarrow \Lambda^{\downarrow}X)}}}
  =
  \frac{d\delta\sigma^{\Lambda}}{d\sigma^{\Lambda}},
  \label{eq:dttDef}
\end{equation}
where $\uparrow$\,($\downarrow$) denotes the positive\,(negative) transverse polarization direction of the particles and $\delta\sigma^{\Lambda}$ is the transversely polarized cross section.
Within the factorization framework,  $\delta\sigma^{\Lambda}$ can be written as the convolution of parton transversity, polarized partonic cross section and the polarized fragmentation function\,\cite{deFlorian:1998am}.

The polarization of $\Lambda\,(\bar{\Lambda})$ hyperons, $P_{\Lambda\,(\bar{\Lambda})}$, can be measured from the angular distribution of the final state particles via their weak decay channel $\Lambda\rightarrow p\pi^{-}\,(\bar{\Lambda}\rightarrow \bar{p}\pi^{+})$,
\begin{equation}
  \frac
  {dN}{\,d\cos{\theta^{*}}\,}
  \propto
  A
  \left(1+\alpha_{\Lambda\,(\bar{\Lambda})}P_{\Lambda\,(\bar{\Lambda})}\cos{\theta^{*}}\right),
  \label{eq:distFinalState}
\end{equation}
where $A$ is the detector acceptance varying with $\theta^{*}$ as well as other observables, $\alpha_{\Lambda\,(\bar{\Lambda})}$ is the weak decay parameter, and $\theta^{*}$ is the angle between the $\Lambda\,(\bar{\Lambda})$ polarization direction and the (anti-)\,proton momentum in the $\Lambda\,(\bar{\Lambda})$ rest frame. 
For the $D_{\mathrm{TT}}$ measurements, the transverse polarization direction of the outgoing fragmenting parton is used to obtain $\theta^{*}$.
Since there is a rotation along the normal direction of the scattering plane between the transverse polarization directions of incoming and outgoing quarks\,\cite{JetPolar:1993kq}, the momentum direction of the outgoing parton is required and the reconstructed jet axis adjacent to the $\Lambda\,(\bar{\Lambda})$ is used as the substitute for the direction of the outgoing fragmenting quark.
Previously, only the spin transfer along the normal direction of the $\Lambda$ production plane, $D_{\mathrm{NN}}$, was measured with fixed target proton-proton collisions by the E704 Collaboration at Fermilab\,\cite{Bravar:1997fb}.

The data for the $D_{\mathrm{TT}}$ measurement here were collected at RHIC with the STAR experiment in the year 2012 with transversely polarized proton beams, which amount to an integrated luminosity of $18\,\mathrm{pb}^{-1}$. 
The average transverse polarizations for the two beams were $58\%$ and $64\%$.
%
The primary detector sub-system is the Time Projection Chamber (TPC)\,\cite{TPC}, which provides tracking for charged particles in the pseudo-rapidity range of {\color{blue}$\left|\eta\right|<1.3$} with full azimuthal coverage.
The Barrel Electromagnetic Calorimeter\,(BEMC)\,\cite{BEMC} and Endcap Electromagnetic Calorimeter\,(EEMC)\,\cite{EEMC} were used in generating the pimary jet trigger information at STAR.
The data sample for this analysis were recorded with jet-patch\,(JP) trigger conditions which required a transverse energy deposit $E_{\mathrm{T}}$ in BEMC or EEMC patches (each covering $\Delta\eta\times\Delta\phi = 1\times1$) exceeding certain thresholds.

The $\Lambda$ and $\bar{\Lambda}$ candidates were reconstructed from their dominant weak decay channels, $\Lambda\rightarrow p\pi^{-}$ and $\bar{\Lambda}\rightarrow \bar{p}\pi^{+}$, with a branching ratio of $63.9\%$\,\cite{PDG}.
The selection procedure of $\Lambda$ and $\bar{\Lambda}$ candidates is based on the topology of the weak decay in a similar way as the previous longitudinal spin transfer measurement  in Ref.\,\cite{Abelev:2009xg}.
In addition, reconstructed jets were employed to obtain the momentum direction of the fragmenting quark and thus the transverse polarization direction of hyperons for the $D_{\mathrm{TT}}$ measurement.
Jets were reconstructed using the anti-$k_{\mathrm{T}}$ algorithm\,\cite{antiKt:2008gp} and
then the association between $\Lambda\,(\bar{\Lambda})$ candidates and the adjacent reconstructed jet was made by constraining the radius,
$\Delta R=\sqrt{(\Delta\eta)^{2}+(\Delta\phi)^{2}}<0.6\,$,
between the $\Lambda\,(\bar{\Lambda})$ momentum direction and the jet axis in $\eta-\phi$ space.
The bin counting method was used to obtain the raw yields of $\Lambda$ and $\bar{\Lambda}$ candidates. 
The signal mass windows were chosen to be about $3\sigma$ of the mass width.
The residual background fractions $r$ were estimated by the side-band method and found to be $7\%\sim10\%$.
The background subtraction was done at the spin transfer level as $D_{\mathrm{TT}} =
  (\,D_{\mathrm{TT}}^{\mathrm{raw}}-rD_{\mathrm{TT}}^{\mathrm{bkg}}\,)/(1-r)$, with $D_{\mathrm{TT}}^{\mathrm{raw}}$ and $D_{\mathrm{TT}}^{\mathrm{bkg}}$ defined as the spin transfer for signal and side-band regions.

To minimize the systematics associated with acceptance and relative luminosity, $D_{\mathrm{TT}}$ for hyperon candidates has been extracted from the asymmetry in small $\cos{\theta^{*}}$ intervals using\cite{Adam:2018dtt}
\begin{equation}
  D_{\mathrm{TT}} = 
  \frac{1}{\,\alpha P_{\mathrm{beam}} \left<\cos{\theta^{*}}\right>\,}
  \frac
    {\,\sqrt{N^{\uparrow}(\cos{\theta^{*}})N^{\downarrow}(-\cos{\theta^{*}})\,}
  -\sqrt{N^{\downarrow}(\cos{\theta^{*}})N^{\uparrow}(-\cos{\theta^{*}})\,}\, }
  {\,\sqrt{N^{\uparrow}(\cos{\theta^{*}})N^{\downarrow}(-\cos{\theta^{*}})\,}
  +\sqrt{N^{\downarrow}(\cos{\theta^{*}})N^{\uparrow}(-\cos{\theta^{*}})\,}\, },
  \label{eq:dttAsy}
\end{equation}
where $\alpha_{\Lambda}=0.642\pm0.013$~\cite{PDG}, $\alpha_{\bar{\Lambda}} = -\alpha_{\Lambda}$, $P_\mathrm{beam}$ is the beam polarization and $\left\langle\cos{\theta^{*}}\right\rangle$ denotes the average value in the $\cos{\theta^{*}}$ interval.
$N^{\uparrow(\downarrow)}$ is the $\Lambda$ yield in the corresponding $\cos\theta^*$ bin when the proton beam is upward (downward) polarized.
The relative luminosity between $N^{\uparrow}$ and $N^{\downarrow}$ is cancelled in this cross-ratio asymmetry. 
The acceptance is also cancelled as the acceptance in a small $\cos\theta^*$ interval is expected to remain the same when flipping the beam polarization\,\cite{Abelev:2009xg}.

\begin{figure}[htbp]
  \centering
  \includegraphics[width=0.49\textwidth]{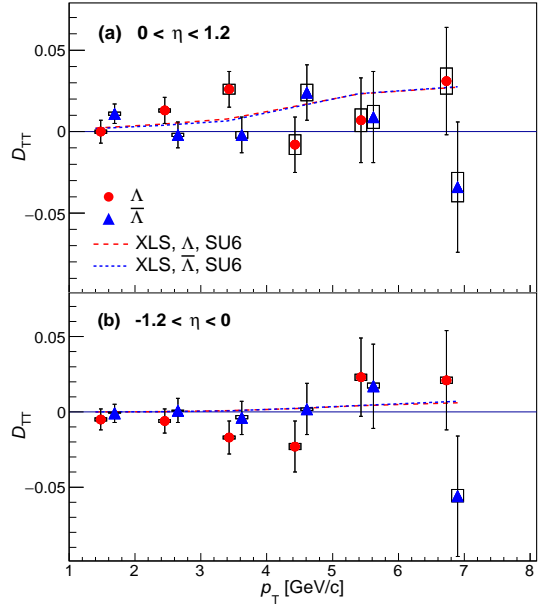}
  \caption{
  The spin transfer $D_\mathrm{TT}$ for $\Lambda$ and $\bar{\Lambda}$ versus $p_\mathrm{T}$ in polarized proton-proton collisions at $\sqrt{s}=200\,\mathrm{GeV}$ at STAR \cite{Adam:2018dtt}, in comparison with model predictions\,\cite{Xu:2004es,Xu:2005ru} for (a) positive $\eta$ and (b) negative $\eta$.
  The vertical bars and hollow rectangles indicate the sizes of the statistical and systematic uncertainties, respectively. The $\bar{\Lambda}$ results have been offset to slightly larger $p_{\mathrm{T}}$ values for clarity.
  }
  \label{fig:dttPt}
\end{figure}

The STAR results for $D_\mathrm{TT}$ versus hyperon $p_\mathrm{T}$ are shown in Fig.\,\ref{fig:dttPt} for $\Lambda$ and $\bar{\Lambda}$ at both positive and negative $\eta$ regions relative to the polarized beam in proton-proton collisions at $\sqrt{s}=200\,\mathrm{GeV}$. 
These results provide the first measurements on transverse spin transfer of hyperons in proton-proton collisions at a high energy of $\sqrt{s}=200$ GeV.
The $D_{\mathrm{TT}}$ results for $\Lambda$ and $\bar{\Lambda}$ are consistent with zero within uncertainties. 
The data cover $p_\mathrm{T}$ up to $7\,\mathrm{GeV}/c$, where $D_{\mathrm{TT}} = 0.031 \pm 0.033(\mathrm{stat.}) \pm 0.008(\mathrm{sys.})$ for $\Lambda$ and $D_{\mathrm{TT}} = -0.034 \pm 0.040(\mathrm{stat.}) \pm 0.009(\mathrm{sys.})$ for $\bar{\Lambda}$ at $\left<\eta\right> = 0.5$ and $\left<p_{\mathrm{T}}\right> = 6.7\,\mathrm{GeV}/c$. 
The data are comparable to a model estimation calculated at $\left<\eta\right>=\pm 0.5$, which is available for RHIC energy, with simple assumptions of the transversity (using the DSSV helicity distribution as input) and the SU6 picture of fragmentation functions\,\cite{Xu:2004es,Xu:2005ru}.
Since strange and anti-strange quarks are expected to carry a significant part of the spins of $\Lambda$ and $\bar{\Lambda}$, the measurements of transverse spin transfer to them can provide sensitivity to the transversity distribution of strange quarks.
Knowledge on transversity of valence quarks has been learned mostly from DIS experiments,
but the transversity distribution of (anti-)strange quarks is still poorly known in experiment\,\cite{Anselmino:2013vqa,Kang:2015msa,Radici:2018iag}.

\section{Improved measurement of longitudinal spin transfer to $\Lambda$ ($\bar{\Lambda}$) hyperons at STAR}

The longitudinal spin transfer, $D_{LL}$, to $\Lambda$  hyperon in polarized proton-proton collisions is defined as:
\begin{equation}
D_{LL}\equiv \frac
{d\sigma_{p^+p \to  \Lambda ^+ X}-d\sigma_{p^+p \to  \Lambda ^-X}}
{d\sigma_{p^+p \to  \Lambda ^+ X}+d\sigma_{p^+p \to  \Lambda ^-X}},
\label{gener1}
\end{equation}
where the superscripts $+$ or $-$ denote the helicity of the beam proton or the $\Lambda$ hyperon. 

The data sample for this analysis was recorded with the STAR experiment in 2009 and corresponds to an integrated luminosity of about 19\,pb$^{-1}$ with an average longitudinal beam polarization, $P_\mathrm{beam}$, of 57\%$\pm$ 4.7\%.
The figure-of-merit of this data sample, $P_\mathrm{beam}^2\mathcal{L}$, is about twelve times higher than that of the previous STAR measurement of $D_{LL}$~\cite{Abelev:2009xg}.
Similar to the 2012 data sample, the events here were recorded also with a jet-patch trigger condition in the BEMC or EEMC.

The reconstruction of $\Lambda\,(\bar{\Lambda})$ is similar to that done for the $D_{TT}$ analysis. 
The detailed selection cuts were tuned in $p_T$ intervals so as to keep as much signal as possible, while keeping the residual background at an acceptable level of about 10\%.
In addition, the $\Lambda$ and $\bar{\Lambda}$ baryons were required to be associated with a reconstructed jet that satisfied the trigger conditions.
Here, the jet sample was reconstructed using the mid-point cone algorithm, as in several previous STAR jet analyses~\cite{Abelev:2007vt,Adamczyk:2012qj}.
For this association, no significant difference is expected either for anti-kt or mid-point cone jet algorithm.
The association required that the reconstructed $\Lambda$ or $\bar{\Lambda}$ candidate was within the jet cone of radius $\Delta {R} =\sqrt{(\Delta\eta)^2+(\Delta \phi)^2} <0.7$.  

Following our previous $D_{LL}$ measurement~\cite{Abelev:2009xg}, the longitudinal spin transfer $D_{LL}$ was extracted from the asymmetry:
\begin{equation}
D_{LL}=\frac{1}{\alpha_{\Lambda(\bar{\Lambda})} P_\mathrm{beam} \left<\cos \theta^*\right>} \frac{N^+ -R N^-} {N^+ + R N^-},
\label{eq_dll}
\end{equation}
where $N^+$ $(N^-)$ is the number of $\Lambda$ or $\bar{\Lambda}$ candidates in the $\cos\theta^*$ interval when the beam is positively (negatively) polarized.
$\left<\cos\theta^*\right>$ is the average value of $\cos\theta^*$ in this interval, but different from the case for $D_{TT}$, here $\theta^*$ is defined as the angle between $\Lambda\,(\bar{\Lambda})$ momentum direction and the (anti-)\,proton momentum in the $\Lambda\,(\bar{\Lambda})$ rest frame.
$R$ denotes the relative luminosity ratio for these two beam polarization states and has been measured with Beam-Beam Counters~\cite{Abelev:2009xg}.

%

\begin{figure}
\begin{center}
\includegraphics[width=0.58\textwidth]{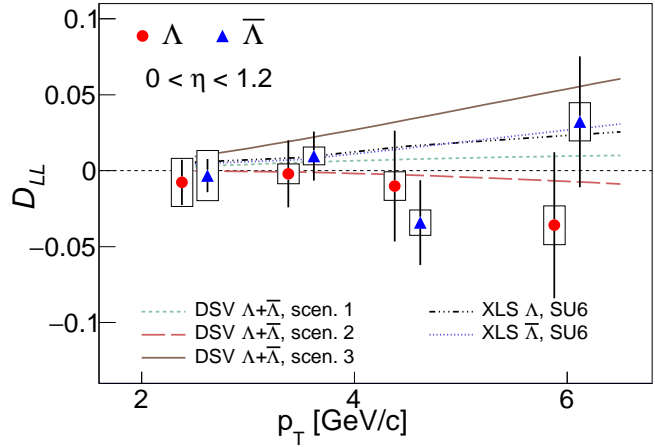}
\caption{Comparison of the measured spin transfer $D_{LL}$ from STAR 2009 data \cite{Adam:2018kzl} with theory predictions for positive $\eta$ versus hyperon $p_T$. The vertical bars and boxes indicate the sizes of the statistical and systematic uncertainties, respectively. The $\bar{\Lambda}$ results have been offset to slightly larger $p_T$ values for clarity.}
\label{DLL_09etap}
\end{center}
\end{figure}

Figure~\ref{DLL_09etap} shows our new STAR results on the spin transfer $D_{LL}$ to the $\Lambda$ and to the $\bar{\Lambda}$ as functions of $p_T$ for positive $\eta$.
The longitudinal spin transfer is found to be $D_{LL}$ = -0.036 $\pm$ 0.048\,(stat) $\pm$ 0.013\,(sys) for $\Lambda$ and $D_{LL}$ = 0.032 $\pm$ 0.043\,(stat) $\pm$ 0.013\,(sys) for $\bar{\Lambda}$ at $\left<\eta\right>$ = 0.5 and $\left<p_T\right>$ = 5.9\,GeV/$c$.
The data are compared with the theory expectations from Ref.~\cite{deFlorian:1998ba}, with $D_{LL}$ for $\Lambda$ and $\bar\Lambda$ combined, and from Ref.~\cite{Xu:2005ru}, with $D_{LL}$ separately for $\Lambda$ and for $\bar\Lambda$.
The data are still consistent with zero within uncertainty, and also provide no evidence for a difference between  $\Lambda$ and $\bar{\Lambda}$ $D_{LL}$.
However, the data are below the DSV scenario 3 theory expectation assuming that the quark polarized fragmentation functions are flavor-independent~\cite{deFlorian:1998ba}.

\section{Summary and Outlook}
In summary, we report the first measurement of the transverse spin transfer, $D_{\mathrm{TT}}$, to $\Lambda$ and $\bar{\Lambda}$ in transversely polarized proton-proton collisions at $\sqrt s$\,=\,200\,GeV at RHIC-STAR, and an improved measurement of the longitudinal spin transfer, $D_{LL}$, to $\Lambda$ hyperons and $\bar{\Lambda}$ anti-hyperons in longitudinally polarized proton--proton collisions at $\sqrt{s}$ = 200\,GeV from STAR.
The results for $D_\mathrm{TT}$ are found to be consistent with zero for $\Lambda$ and $\bar{\Lambda}$ within uncertainties, and are also consistent with model predictions.
The $D_{LL}$ data do not provide evidence for a non-vanishing spin transfer within uncertainties, but tend to be below a model expectation based on the extreme assumption on polarized fragmentation functions.

STAR is expanding its acceptance by series of detector upgrades including an ongoing upgrade to the inner sectors of the TPC and the proposed forward detector upgrades~\cite{starFWD},  which would enable a number of interesting hyperon polarization measurements, including spin transfers $D_{TT}$, $D_{LL}$ and other hyperon polarization measurements such as the induced transverse polarization in unpolarized proton-proton collisions, at very forward rapidities in proton-proton collisions.

\end{document}